\documentclass[fleqn,aip,amsmath,amssymb,reprint]{revtex4-1}

\usepackage{setspace}
\usepackage[utf8]{inputenc}
\newcommand{\R}{{\mathbb{R}}}

\usepackage{xcolor,graphicx}
\usepackage{color}
\usepackage{amsmath}

\usepackage[normalem]{ulem}

\renewcommand{\cite}[1]{[\onlinecite{#1}]}

\begin{document}
\title{Bound states of a one-dimensional Dirac equation with multiple
delta-potentials
}
\author{V. P. Gusynin}
\address
{Bogolyubov Institute for Theoretical Physics, National Academy of
	Sciences of Ukraine, Kyiv 03143, Ukraine}
\author{O. O. Sobol}
\address
{Taras Shevchenko National University of Kyiv, 64/13 Volodymyrska Street, Kyiv 01601, Ukraine}
\author{A. V. Zolotaryuk}
\address
{Bogolyubov Institute for Theoretical Physics, National Academy of
	Sciences of Ukraine, Kyiv 03143, Ukraine}
\author{Y. Zolotaryuk}
\address
{Bogolyubov Institute for Theoretical Physics, National Academy of
	Sciences of Ukraine, Kyiv 03143, Ukraine}

\date{\today}

\begin{abstract}

Two approaches are developed for the study of the bound states of a
one-dimensional Dirac  equation with the potential consisting of $N$
$\delta$-function centers.   One of these uses the Green's function method.
This method is applicable to a finite number $N$ of $\delta$-point
centers, reducing the bound state problem to finding the energy
eigenvalues  from the determinant of a $2N\times2N$ matrix. The second
approach starts with the matrix for a single delta-center that connects the
two-sided boundary conditions for this center. This connection
matrix is  obtained from the squeezing limit
of a piecewise constant approximation of the delta-function. Having then
the connection matrices for each center, the transmission matrix for
the whole system is obtained by multiplying the one-center
connection matrices and the free transfer matrices between neighbor
centers. An equation for bound state energies is derived in terms of
the elements of the total transfer matrix. Within both the approaches,
the transcendental equations for bound state energies are derived, the
solutions to which depend on the strength of delta-centers and the distance between
them, and this dependence is illustrated by numerical calculations.
The bound state energies for the potentials composed of one, two, and three
delta-centers ($N=1,\,2,\,3$) are computed explicitly. The principle
of strength additivity is analyzed in  the limits as
the delta-centers merge at a single point or diverge to infinity.

  \end{abstract}
\maketitle




\section{Introduction}

Point interactions in one spatial dimension are often used to approximate in a simple way
realistic physical systems producing exactly solvable models. In  quantum mechanics,
on the basis of these models, it is possible  to describe very effectively
different phenomena such as scattering, bound states, etc. in terms of
elementary functions. A whole class of these models is represented by
one-dimensional Schr\"{o}dinger and Dirac equations. Intuitively, the point interactions
are understood as sharply localized potentials,   providing  relatively
simple situations, where an appropriate way of squeezing to zero
should  be chosen  in relevance with a real structure (see books \cite{do,a-h}
for details and references therein).

In many cases of quantum mechanics, it is very convenient to approximate
finite-range realistic potentials by Dirac's delta-function $\delta(x)$,
$x$ is the spatial coordinate.
 While there are no problems to treat the one-dimensional Schr\"{o}dinger equation
with one or several $\delta$-potentials,  Sutherland and Mattis \cite{sm},
early in 1981, were surprised  to discover difficulties and ambiguities
regarding the use of the delta-function potential in a one-dimensional Dirac equation.
They   were able to construct two examples with ``initial'' different shapes
that  approximate the function $\delta(x)$, resulting in different values
of scattering coefficients and other properties at a point (say, $x=0$) of discontinuity.
This non-uniqueness problem emerges from the fact that the Dirac equation must be
examined on the wavefunctions (spinors), which are themselves discontinuous at centers of
discontinuity. Indeed, in this case, the Dirac equation contains the products of
distributions, like a delta-function multiplied by a discontinuous wavefunction, generally
being not an algebra.

A similar ambiguity problem has been arisen  two decades later \cite{c-g}
for a one-dimensional Schr\"{o}dinger equation with the potential in the form
of the spatial derivative $\delta'(x)$ of the delta-function. Again, in this case,
the Schr\"{o}dinger equation must be defined on the family of discontinuous functions,
 having the product of distributions at the point of discontinuity. On some particular
examples, including a piecewise constant approximations of the distribution $\delta'(x)$,
it was noticed the dependence of the realized point interactions on the way of
approximation.  Using a general family of all possible approximations to $\delta'(x)$,
Golovaty with coworkers \cite{Golovaty} have  rigorously established this shape-dependence.
They related this phenomenon to the existence of a hidden parameter in the
Schr\"{o}dinger equation with the $\delta'$-potential, contrary to the Schr\"{o}dinger
equation with $\delta$-potential.

Because of ambiguous results and many discussions on relativistic point interactions,
a large body of literature in addition to \cite{sm} (see, e.g., the
papers \cite{Subramanian1971,Lapidus,Calkin,Benvengu1994,hu,bcl,adv,km,ntd}, a few to mention), including the
study \cite{Peter,Guilarte} on scattering properties, is devoted to the considered problem.
A special interest is paid
to relativistic models with two attracting point-like potentials in the form of
double-delta functions (see \cite{Fillion,Gorbar} and references therein)
used for various physical applications. The models with triple-delta
potentials \cite{cs} are also of interest for realizing point interactions.

In the present work, we study the bound states of a one-dimensional Dirac equation with
the potentials, which are composed of one, two, and three delta-function centers.
We utilize two approaches: (i) the Green's function analysis  and (ii)
a piecewise constant approximation of delta-functions. The key point is
the comparison of the limits as the distance between delta-centers tends to zero
or infinity. Note that the  bound state spectrum of the one-dimensional
Schr\"{o}dinger equation with multiple delta-potentials has been examined in the recent work \cite{Erman}.

The one-dimensional Dirac equation with a potential $V(x)$  is generally given by
\begin{equation}
\left[-{\rm i}\alpha {d \over dx} -\beta m + V(x) \right]\psi(x)=E\psi(x),
\label{1}
\end{equation}
where $\psi(x) = \mbox{col}\left(\psi_1(x),\psi_2(x)\right)$ is a spinor, $V(x)$
a scalar potential,
$m$ the fermion mass and $E$  the energy. The two-dimensional matrices
$\alpha$ and $\beta$ can be expressed in terms of the Pauli $\sigma$-matrices.

For a convenience, we use the Majorana representation, where
$\alpha =- \sigma_y \equiv - \sigma_2$ and $\beta =\sigma_z \equiv \sigma_3 $, so that
Eq.\,(\ref{1}) can be rewritten in the form of two coupled equations with real coefficients:
 \begin{equation}
\begin{array}{ll}
\smallskip
 \psi'_1(x) = (m -E +V)\psi_2(x), \\
\psi'_2(x) = (m + E -V)\psi_1(x), \end{array} \label{2}
\end{equation}
 where the prime stands for the differentiation over $x$.

The present paper is an extension of the works \cite{Gorbar,Fillion},
where the bound states of the equation with double-delta potentials were
investigated.   In general, a multiple delta-potential $V(x)$
can be represented as the sum  of $N$ delta-function centers:
\begin{equation}
 V(x) = \sum_{i =1}^N g_i \delta(x - r_i),
 \label{3}
\end{equation}
where each $g_i \in \R$ is a strength of the $i$th $\delta$-interaction
localized at a point $x=r_i \in \R$. In the present work, we restrict ourselves
to the cases with $N=1,\,2,$ and 3.

\section{Bound states of a single-, double-, and triple-delta potentials: The Green's function approach }

We first consider the Lippmann-Schwinger approach which is the most suitable one
for describing electron dynamics of free electrons moving in the field of a finite set of $\delta$-point centers
(\ref{3}). This approach is applicable for describing such dynamics in an arbitrary space dimension and for various
types of electron dispersion. For a given Hamiltonian $H=H_0+V$, where $H_0$ is a free Hamiltonian, we define the Green's operators
\begin{equation}
(H-E)G=1, \quad (H_0-E)G_0=1.
\end{equation}
All the important physical information contains in the total Green's function 
at the energy $E$,
\begin{equation}
G({\bf r},{\bf r}^{\prime}; E)=\langle {\bf r}|G|{\bf r}^{\prime}\rangle=\left 
\langle {\bf r}\left |\frac{1}{H-E}\right|{\bf r}^{\prime}\right \rangle,
\end{equation}
and the free Green's function becomes
\begin{equation}
 G_{0}({\bf r},{\bf r}^{\prime}; E)=\langle {\bf r}|G_0|{\bf r}^{\prime}\rangle
 =\left \langle {\bf r}\left |\frac{1}{H_{0}-E}\right |{\bf r}^{\prime}\right \rangle
\end{equation}
(infinitesimal imaginary part is implied in the energy $E$: $E + i0$).

The Lippmann-Schwinger equation for the Green's function has the form
\begin{eqnarray}
&&G({\bf r},{\bf r}^{\prime}; E) = G_{0}({\bf r},{\bf r}^{\prime}; E)-\nonumber\\
&&-\int d{\bf r}^{\prime\prime}G_{0}({\bf r},{\bf r}^{\prime\prime}; E)
V({\bf r}^{\prime\prime})G({\bf r}^{\prime\prime},{\bf r}^{\prime}; E),
\label{Lippmann-Schwinger}
\end{eqnarray}
that in the case of the $n$-dimensional version of the potential (\ref{3}),
leads to the equation
\begin{equation}
\hspace{-10mm}G({\bf r},{\bf r}^{\prime}; E) = G_{0}({\bf r},{\bf r}^{\prime}; E)-
\sum_{j}G_{0}({\bf r},\mathbf{r}_{j}; E)g_{j} G(\mathbf{r}_{j},{\bf r}^{\prime}; E).\\
\end{equation}
Setting  $\mathbf{r}=\mathbf{r}_{i}$ in this equation, we get the solution
\begin{eqnarray}
&&G({\bf r},{\bf r}^{\prime}; E) = G_{0}({\bf r},{\bf r}^{\prime};  E)-\nonumber\\
&&-\sum_{ij}G_{0}({\bf r},\mathbf{r}_{i};  E)(\Delta^{-1})_{ij}G_{0}
(\mathbf{r}_{j},{\bf r}^{\prime};  E),
\end{eqnarray}
where the matrix $\Delta$ is
\begin{equation}
\Delta_{ij}(E)\equiv \frac{\delta_{ij}}{g_{j}}\mathbf{1}
+G_{0}(\mathbf{r}_{i},\mathbf{r}_{j}; E).
\end{equation}
Note that indices $i$ and $j$ denote the $2\times 2$ blocks of the matrix $\Delta$
and not the single matrix elements. Indeed, $\mathbf{1}$ is the $2\times 2$ unity matrix
and $G_{0}$ is the free Green's function of a one-dimensional Dirac equation
being a $2\times 2$ matrix as well. Thus, for the potential with $N$ delta-functions,
one gets a $2N\times 2N$ matrix $\Delta$.
The discrete levels of the system are the roots of the equation ${\rm det}\,\Delta(E)=0$.
We now apply this approach to the one-dimensional  Dirac equation with the potential (\ref{3}).

The free Green’s function for the one dimensional Dirac equation (\ref{1})
is translationally invariant. It is given by the expression
\begin{equation}
G_0(x-x',E)=\int\limits_{-\infty}^\infty\frac{d p}{2\pi}{\rm e}^{-{\rm i}p(x-x')}
\frac{\sigma_2p-\sigma_3m+E}{p^2+m^2-E^2}\,.
\end{equation}
 For $G_0(x-x',E)$ we get the explicit expression
\begin{eqnarray}
\hspace{-10mm}G_0(x-x',E)&=&\frac{1}{2}\left[-{\rm i}\sigma_2{\rm sign}(x-x')
+\frac{-\sigma_3m+E}{\sqrt{m^2-E^2}}\right]\nonumber\\
&\times& {\rm e}^{-|x-x'|\sqrt{m^2-E^2}}.
\end{eqnarray}
Here ${\rm sign}(x)$ is defined as having values $-1,0,1$ for $x<0$, $x=0$, $x>0$, respectively.
For the coincidence limit of points, we obtain
\begin{equation}
G_0(0,E)=\frac{-\sigma_3m+E}{2\sqrt{m^2-E^2}}\,.
\end{equation}

\subsection{A single-delta potential: $N=1$, $g_1 \equiv -g$, $r_1=0$  }

For a one-delta center, the matrix $\Delta$ has the simple form
\begin{equation}
\Delta=-\frac{1}{g}\mathbf{1}+\frac{-\sigma_3m+E}{2\sqrt{m^2-E^2}}\,,
\end{equation}
hence
\begin{equation}
{\rm det\,}\Delta=\left(\frac{1}{g}+\frac{\rho}{2}\right)\left(\frac{1}{g}-\frac{1}{2\rho}\right),
\end{equation}
where
\begin{equation}
\rho:=\sqrt{\frac{m-E}{m+E}}\,\,.
\label{def:rho}
\end{equation}
A non-trivial solution of the equation ${\rm det\,}\Delta=0$ for $g>0$ occurs only 
if $\rho =g/2$, resulting in
the following expression for the bound state energy:
\begin{equation}
E=m\frac{4-g^2}{4+g^2}, ~~~ |E|<m.
\label{7}
\end{equation}

The bound state appears at any small attractive constant $g$, but its energy never
reaches the edge $E=-m$ of the negative continuum spectrum.

\subsection{A symmetric double-delta potential: $N=2$, $g_1 =g_2 \equiv -g$,
$r_1 = - r_2 \equiv - R$}

For a two-delta symmetric  potential, keeping finite $R$, the matrix $\Delta$ takes 
the form
\begin{eqnarray}
\hspace{-5mm}\Delta=\left(\begin{array}{cccc} \medskip -\frac{1}{g}-\frac{\rho}{2}&0&
-\frac{\rho}{2}{\rm e}^{-2z}&\frac{1}{2}{\rm e}^{-2z}\\ \medskip
0&-\frac{1}{g}+\frac{1}{2\rho}&-\frac{1}{2}{\rm e}^{-2z}&\frac{1}{2\rho}{\rm e}^{-2z}\\
\medskip
-\frac{\rho}{2}{\rm e}^{-2z}&-\frac{1}{2}{\rm e}^{-2z}&-\frac{1}{g}-\frac{\rho}{2}&0\\
\frac{1}{2}{\rm e}^{-2z}& \frac{1}{2\rho}{\rm e}^{-2z}&0&-\frac{1}{g}+\frac{1}{2\rho}
\end{array}\right),
\label{matrix:double-delta}
\end{eqnarray}
where $z :=R\sqrt{m^2-E^2}$.

The determinant of this matrix reads
$$
{\rm det\,}\Delta=\frac{(2\rho-g)^2(2+g\rho)^2-4g^2{\rm e}^{-4z}(1+\rho^{2})^2}{16g^4\rho^{2}}.
$$
Introducing the new variable
\begin{equation}
	\label{var-X}
	X=\frac{E}{\sqrt{m^{2}-E^{2}}} \frac{g}{1-g^2/4},
\end{equation}
the equation ${\rm det\,}\Delta=0$ takes the form
\begin{equation}
	\label{eq-2s-X}
	\!\!\!\!\!\!\!\!(1-{\rm e}^{-4z})X^2-2X+1-{\rm e}^{-4z}
	\frac{g^{2}}{(1-g^2/4)^{2}}=0.
\end{equation}
This quadratic equation can formally be solved, leading to the following 
two equations for bound state energies:
\begin{equation}
\!\!\!\!\!\!\!\!\!\!\!\!\!
\left( 1- {g^2 \over 4}\right)\sqrt{m^2-E^2}=g\left[E\pm m
{\rm e}^{-2R\sqrt{m^2-E^2}}\right].~~
\label{12}
\end{equation}
Equation (\ref{12}) coincides exactly with Eq.\,(56) in \cite{Fillion}
and the cases $\pm$ correspond to the ground  and first excited states, respectively.
In the limit as $R \to \infty$,  the solution of Eq.\,(\ref{12}) gives the energy
for a single center given by Eq.\,(\ref{7}).
However, as $R \to 0$, Eq.\,(\ref{12}) has a solution
$$
E=m\frac{\left(g^4-24 g^2+16\right)}{\left(g^2+4\right)^2}\mathrm{sign}\left(4-g^2\right),
$$
which does not reproduce the energy for one center
with the charge $2g$ as expected.

\subsection{A dipole-like delta-potential:  $N=2$, $g_1 = - g_2 \equiv -g$,
$r_1 = - r_2 \equiv - R$}

For an antisymmetric double-delta potential the matrix $\Delta$ has again the form (\ref{matrix:double-delta})
where $g$ is replaced with $-g$ in the $2\times2$ matrix in the lower right corner.
The determinant of this matrix equals
\begin{equation*}
{\rm det\,}\Delta =\frac{(4\rho^2-g^2)(4-g^2\rho^2)+4g^2 {\rm e}^{-4z}(1+\rho^2)^2}{16g^4\rho^2}.
\end{equation*}
For $R=0$, this determinant  reduces to ${\rm det\,}\Delta =(4+g^2) /16g^4$\,, that means non-existence of bound states. In the other limit as  $z \to \infty$,
we have
\begin{equation}
{\rm det\,}\Delta=\frac{(4-g^2\rho^2)(4\rho^2-g^2)}{16g^4\rho^2}\,,
\label{17}
\end{equation}
hence, from the equation $\det\Delta=0$, we get the bound state energy for one center, i.e.,
\begin{equation}
E=m\frac{4-g^2}{4+g^2} \quad \mbox{or}\quad E=-m\frac{4-g^2}{4+g^2}\,.
\label{18}
\end{equation}
These two levels cross at $g=2$ with $E=0$.

For finite $R$, the bound state energies are determined from the equation ${\rm det\,}\Delta=0$, which can be rewritten in terms of a new variable $X$ [defined in Eq.~(\ref{var-X})] as follows:
\begin{equation}
	\label{19}
	X^2=\frac{(g^2-4)^2+16g^2{\rm e}^{-4z}}{(g^2-4)^2(1-{\rm e}^{-4z})}.
\end{equation}
This equation has the symmetry under the change $X\to -X$, meaning that if we have the solution with an energy $E$, then there is also the solution with $-E$.
Equation (\ref{19}) can also be rewritten in the form
\begin{equation}
E=\pm \,m\sqrt{1-(1-{\rm e}^{-4z})\frac{16g^2}{(g^2+4)^2}}.
\label{20}
\end{equation}
As follows from these equations, the bound state  energies are monotonic functions of
the distance $R$.

\subsection{A three-delta potential of the same polarity: $N=3$, $g_1 =g_2=g_3 
\equiv -g$, $r_1 \equiv - R_1$, $r_2 \equiv R_2$, $r_3 =0$}
\label{subsec-symmetric-triple-delta-Green}

Let us consider now the case of a potential containing three delta-functions 
of the same strength. In this case, the matrix $\Delta$ is $6\times 6$, it can be constructed in the same way as in Eq.~(\ref{matrix:double-delta}) and has the form
\begin{widetext}
	\begin{equation}
		\label{Delta-3s}
		\Delta=\left(
		\begin{array}{cccccc}
			\medskip -\frac{1}{g}-\frac{\rho}{2}&0&
			-\frac{\rho}{2}{\rm e}^{-z_1-z_2}&\frac{1}{2}{\rm e}^{-z_1-z_2} & -\frac{\rho}{2}{\rm e}^{-z_1}&\frac{1}{2}{\rm e}^{-z_1}\\ \medskip
	0&-\frac{1}{g}+\frac{1}{2\rho}&-\frac{1}{2}{\rm e}^{-z_1-z_2}&\frac{1}{2\rho}
	{\rm e}^{-z_1-z_2}&-\frac{1}{2}{\rm e}^{-z_1}&\frac{1}{2\rho}{\rm e}^{-z_1}\\
			\medskip
-\frac{\rho}{2}{\rm e}^{-z_1-z_2}&-\frac{1}{2}{\rm e}^{-z_1-z_2}&-\frac{1}{g}-\frac{\rho}{2}&0&-\frac{\rho}{2}{\rm e}^{-z_2}&-\frac{1}{2}{\rm e}^{-z_2}\\ \medskip			\frac{1}{2}{\rm e}^{-z_1-z_2}& \frac{1}{2\rho}{\rm e}^{-z_1-z_2}&0&-\frac{1}{g}+\frac{1}{2\rho}&\frac{1}{2}{\rm e}^{-z_2}& \frac{1}{2\rho}{\rm e}^{-z_2}\\
			\medskip
-\frac{\rho}{2}{\rm e}^{-z_1}&-\frac{1}{2}{\rm e}^{-z_1}&-\frac{\rho}{2}{\rm e}^{-z_2}&\frac{1}{2}{\rm e}^{-z_2}&-\frac{1}{g}-\frac{\rho}{2}&0\\ \medskip
\frac{1}{2}{\rm e}^{-z_1}&\frac{1}{2\rho}{\rm e}^{-z_1}&-\frac{1}{2}{\rm e}^{-z_2}&\frac{1}{2\rho}{\rm e}^{-z_2}&0&-\frac{1}{g}+\frac{1}{2\rho}
		\end{array}
		\right).
	\end{equation}
\end{widetext}
For its determinant  we obtain
\begin{eqnarray}
\hspace*{-0.8cm}{\rm det\,}\Delta&=&\frac{1}{64g^{6}\rho^{3}}\left[(2\rho-g)^{3}
(2+\rho g)^{3}\right.\nonumber\\
	&-&\left.4({\rm e}^{-2z_1}+{\rm e}^{-2z_2})g^{2}(2\rho-g)(2+\rho g)
	(1+\rho^{2})^{2}\right.\nonumber\\
	&-&\left.4{\rm e}^{-2(z_1+z_2)}g^{2}(2\rho+g)(2-\rho g)(1+\rho^{2})^{2}\right],
\end{eqnarray}
where $z_j=R_j\sqrt{m^2-E^2}$. In terms of the variable $X$,
 given by Eq.~(\ref{var-X}),
the equation ${\rm det\,}\Delta=0$ can be written in the following form:
\begin{equation}
	\label{eq-3s-X}
	G_3 X^3+G_2 X^2+G_1 X+G_0=0,
\end{equation}
where
\begin{equation}
	\begin{array}{llll}
		\smallskip
		G_0 := -1+\frac{g^{2}}{(1-g^2/4)^{2}}\left[ {\rm e}^{-2z_1} +
		 {\rm e}^{-2z_2}+ {\rm e}^{-2(z_1 +z_2)}
		\right] , \\
		\smallskip
		G_1 := 3 - \frac{g^{2}}{(1-g^2/4)^{2}}\left[ {\rm e}^{-2z_1} + 
		{\rm e}^{-2z_2}-
		{\rm e}^{-2 (z_1 +z_2)} \right],\\
		\smallskip
		G_2 := \left( 1+ {\rm e}^{-2z_1}\right) \left(1 + {\rm e}^{-2z_2} \right)-4,\\
		G_3: = \left( 1- {\rm e}^{-2z_1}\right) \left(1 - {\rm e}^{-2z_2} \right).
	\end{array}
	\nonumber
\end{equation}
Numerical solutions of Eq.\,(\ref{eq-3s-X}) for a symmetric configuration of delta-potentials $R_{1}=R_{2}\equiv R$ are shown in Fig.\,\ref{fig-triple-plus} 
by the dashed (red) lines.

Let us study some limiting cases. If we set $R_{2}\to \infty$ keeping $R_{1}\equiv R$ finite, Eq.\,(\ref{eq-3s-X}) takes the form
\begin{equation*}
	\!\!\!\!\!\!\!\!\!\!\!\!\!\!\!\!
	(X-1)\Big[(1-{\rm e}^{-2z})X^2-2X+1-\frac{g^{2}}{(1-g^2/4)^{2}}{\rm e}^{-2z}\Big]=0.
\end{equation*}
Equating the first bracket to zero, we immediately get the solution (\ref{7}) for a single-delta potential. The second bracket coincides with Eq.\,(\ref{eq-2s-X}) with the replacement $2z\to z$, which determines the bound states in a symmetric double-delta potential.

Further, let us set both $R_{1}$ and $R_{2}$ to zero. In this limit,
 Eq.\,(\ref{eq-3s-X}) reduces to
$$
X=\frac{g^4-56 g^2+16}{3 g^4-40 g^2+48},
$$
which gives for the energy eigenvalue
\begin{eqnarray}
\hspace*{-0.85cm}E&=&m\frac{-g^6+60 g^4-240 g^2+64}{\left(g^2+4\right)^3}
{\rm sign}(3 g^4-40 g^2+48)\nonumber\\
\hspace*{-0.85cm}&=&m\cos(3\Theta){\rm sign}(\sin(3\Theta)),
\end{eqnarray}
where $\Theta=2\,{\rm arctan}(g/2)$. This result does not coincide with Eq.\,(\ref{7}) with replacement $g\to 3g$, meaning that there is no additivity of the potential strength. This feature is clearly seen in Fig.\,\ref{fig-triple-plus}(b) where the red dashed curve at $mR\to 0$ does not hit the red dot which denotes the energy level in a single delta-potential with strength $3g$.

Finally, let us set both distances $R_1$ and $R_{2}$ to infinity. Then, 
Eq.\,(\ref{eq-3s-X}) becomes
\begin{equation}
	(X-1)^{3}=0
\end{equation}
giving three coinciding energy levels in a single-delta potential
described by Eq.\,(\ref{7}). 
This behavior is clearly seen in Fig.\,\ref{fig-triple-plus}(b).

\subsection{A three-delta potential with the middle center of opposite polarity:
	$N=3$, $g_1 =g_2= - g_3 \equiv - g$, $r_1 \equiv - R_1$, $r_2 \equiv R_2$, 
	$r_3 =0$}
\label{subsec-alternating-triple-delta-Green}

The matrix $\Delta$ for alternating-sign triple-delta potential has the same form 
as in Sec.\,\ref{subsec-symmetric-triple-delta-Green} with the replacement $g\to -g$ 
in the last two lines. Its determinant has the form
\begin{multline}
	\label{det-3a}
\hspace*{-0.8cm}{\rm det\,}\Delta=\frac{(2\rho-g)(2+\rho g)}{4g^2 \rho}\times\frac{1}{16g^{4}\rho^2}\big\{(4\rho^2-g^2)(4-g^2\rho^2)\\
	+4g^2 (1+\rho^2)^2[{\rm e}^{-2z_1}+{\rm e}^{-2z_2}-{\rm e}^{-2(z_1+z_2)}]\big\}.
\end{multline}
The equation for bound states ${\rm det\,}\Delta=0$ written in terms of the variable 
$X$ reads as
\begin{equation}
	\label{eq-3a-X}
	(X-1)(X^{2}-D)=0,
\end{equation}
where
\begin{equation*}
	D:=\frac{(g^2-4)^2+16g^2({\rm e}^{-2z_1}+{\rm e}^{-2z_2}+{\rm e}^{-2z_1-2z_2})}{(g^2-4)^2(1-{\rm e}^{-2z_1})(1-{\rm e}^{-2z_1})}.
\end{equation*}
The first bracket in Eq.\,(\ref{eq-3a-X}) gives the energy level in a single-delta potential (\ref{7}), while the second bracket gives two levels located symmetrically with respect to zero energy. They can be found from the following equation:
\begin{equation}
	\label{energy-3a}
	\hspace*{-0.8cm}E=\pm m\sqrt{1-\left(1-{\rm e}^{-2z_1}\right) \left(1-{\rm e}^{-2z_2}\right)\frac{16 g^2}{\left(g^2+4\right)^2}}.
\end{equation}
The splitting of the energies of one negatively charged center is physically explained by the fact that it is in the field 
of a dipole with zero total electric charge. 
Numerical solutions to Eq.\,(\ref{eq-3a-X}) are shown by the dashed lines in Fig.~\ref{fig-triple-alt}. The orange line corresponds to a decoupled branch of the spectrum given by Eq.\,(\ref{7}) while the red lines show the solutions of 
Eq.\,(\ref{energy-3a}).

Let us now consider a few limiting cases. If $R_{1}\to \infty$ ($z_1\to\infty$) and $R_{2}\equiv R$ remains constant, Eq.~(\ref{energy-3a}) reduces to Eq.\,(\ref{20}) for the bound states in the dipole potential. If $R_{1}\to 0$ or $R_2\to 0$, two delta functions of the opposite charge annihilate and Eq.\,(\ref{energy-3a}) implies 
$E=\pm m$, i.e., we are left with energy levels of free fermions. 
Indeed, the red dashed lines in Fig.\,\ref{fig-triple-alt}(b) merge with the upper 
and lower continua when $R_1=R_2\to 0$. Finally, if both $R_{1}$ and $R_{2}$ tend 
to infinity, Eq.\,(\ref{energy-3a}) gives $E=\pm m (4-g^2)/(4+g^2)$, i.e., 
the energy levels of single-delta potentials of the opposite charge.

\section{Bound states of a single-, double-, and triple-delta potentials:
A piecewise constant approximation}

Consider the Dirac equation in the same representation given by Eqs.\,(\ref{1})
and (\ref{2}), but now on a finite interval  $x_1 \le x \le x_2$, where
the potential   $V(x)$ is a non-zero constant and outside this interval it is
identically zero. Then, within this interval, the general solution of
 Eqs.\,(\ref{2})  reads
\begin{equation}
\begin{array}{ll}
\medskip
\psi_1(x)  = C_1 {\rm e}^{{\rm i}\gamma x} +C_2  {\rm e}^{ -\,{\rm i}\gamma x}, \\
\psi_2(x) = \eta^{-1} \left(- C_1 {\rm e}^{{\rm i}\gamma x} +
C_2 {\rm e}^{-\,{\rm i}\gamma x} \right), \end{array}
\label{21}
\end{equation}
where
\begin{equation}
\!\!\!\!\!\!\!\!\!\!
 \gamma : = \sqrt{(V - E)^2 -m^2}\,,~~
 \eta :={ \sqrt{m - E +V}  \over \sqrt{m + E- V} }\,,
\label{22}
\end{equation}
so that $\eta|_{V \equiv 0}= \rho$, with $\rho$ defined by Eq.\,(\ref{def:rho}).

Since we are going to treat the Dirac equation on a finite interval
$x_1 \le x \le x_2$, we
need the transfer matrix, that connects the boundary conditions at the
boundaries $x=x_1$ and $x=x_2$. To this end,  we express
the constants $C_1$ and $C_2$ in terms of the values $\psi_1(x_1)$ 
and $\psi_2(x_1)$, and
then use these expressions in the formulas for $\psi_1(x_2)$ and $\psi_2(x_2)$.
As a result, we find the matrix equation
\begin{equation}
\!\!\!\!\!\!\!\!\!\!\!\!\!\!\!
 \left( \begin{array}{ll} \psi_1(x_2) \\ \psi_2(x_2) \end{array} \right)=
 \Lambda \left( \begin{array}{ll} \psi_1(x_1) \\ \psi_2(x_1) \end{array} \right),
 ~~~\Lambda \equiv \left( \begin{array}{ll} \lambda_{11}~\lambda_{22} \\
        \lambda_{21}~\lambda_{22}   \end{array} \right),
 \label{23}
\end{equation}
 where the $\Lambda$-matrix reads
\begin{equation}
\!\!\!\!\!\!\!\!\!\!\!\!
 \Lambda = \left( \begin{array}{lr} \smallskip \cos(\gamma l)~ ~~~-{\rm i}\eta \,
 \sin(\gamma l) \\
 - {\rm i}\eta^{-1}\sin(\gamma l)~ ~\cos(\gamma l) \end{array} \right), 
 ~~~l:= x_2 - x_1.~
 \label{24}
\end{equation}

\subsection{A general equation for bound states}

It is convenient to use an equation for bound states, which can be expressed
 in terms of the elements of
the $\Lambda$-matrix. Since beyond the interval $x_1 \le x \le x_2$, we have
$V \equiv 0$, solution (\ref{21}) is transformed to
\begin{equation}
\!\!\!\!\!\!\!\!\!\!\!\!\!\!
\begin{array}{ll}
\smallskip
 \psi_1(x) = \left\{ \begin{array}{ll} C_2 {\rm e}^{\sqrt{m^2- E^2}\,(x-x_1)}
                      &~~~~~ \mbox{for}~ x < x_1, \\
 C_1 {\rm e}^{- \sqrt{m^2- E^2}\,(x-x_2)}
                      &~~~~~ \mbox{for}~  x >x_2,
                     \end{array} \right. \\
 \psi_2(x)  = {1 \over \rho}
 \left\{ \begin{array}{ll}~~ C_2 {\rm e}^{\sqrt{m^2- E^2}\,(x-x_1)}
                      & \mbox{for}~  x < x_1, \\ \smallskip
-\, C_1 {\rm e}^{- \sqrt{m^2- E^2}\,(x-x_2)}
                      & \mbox{for}~ x> x_2 .
                     \end{array} \right. \end{array}
                     \label{25}
\end{equation}
Inserting next the values obtained from Eqs.\,(\ref{25}) at $x =x_1$ and $x=x_2$ into Eq.\,(23),
we get the equation for bound states expressed in terms of the elements of the $\Lambda$-matrix in the form
\begin{equation}
 \lambda_{12} \rho^{-2} +(\lambda_{11} + \lambda_{22}) \rho^{-1} + \lambda_{21} =0.
 \label{26}
\end{equation}
The similar equation takes place for the bound states of a Schr\"{o}dinger equation
in one dimension  \cite{zz14}.

\subsection{A single delta-potential: $N=1$, $g_1 \equiv -g$, $r_1=0$ }
\label{subsec-single-delta}

Consider a  finite-range  potential
\begin{equation}
V_l(x) = -\, {g \over l}\left\{ \begin{array}{ll} 1 & \mbox{for}~~-\, 
l/2 \le x \le l/2,\\
                               0 & \mbox{otherwise},       \end{array} \right.
\label{27}
\end{equation}
that converges to $V(x) =- g\delta(x)$, $g \in \R$, in the sense of distributions.
Then, in Eq.\,(\ref{24}),  $\gamma l \to |g|$ and
$\eta \to {\rm sign}(g) {\rm i}$, so that in the limit as $l \to 0$,
the $\Lambda$-matrix for  one $\delta$-center becomes
\begin{equation}
 \Lambda_\delta = \left(\begin{array}{ll} \cos g ~-\sin g \\
                         \sin g~~~~ \cos g \end{array} \right),~~~g \in \R.
 \label{28}
\end{equation}
Such a matrix was also obtained in the paper \cite{mcks87prc}.
Inserting these matrix elements into Eq.\,(\ref{26}), we get the  equation
\begin{equation}
\rho^{-2} - 2\cot g\cdot \rho^{-1} -1 =0,~~~g \in \R,
\label{29}
\end{equation}
or
\begin{equation}
\frac{\sqrt{m^2-E^2}}{E}=\tan g.
\label{30}
\end{equation}
Hence, the bound state energy for a single delta-potential reads
\begin{equation}
E =  m\, \cos(g)\, {\rm sign}(\sin g), ~~~g \in \R,
 \label{31}
\end{equation}
where the sign-function is defined as ${\rm sign}(x)=1$ for $x\ge 0$ and 
${\rm sign}(x)=-1$ for $x<0$.
 For small $g$, the expression (\ref{31}) coincides with the relation (\ref{7}).
 We note, however, that the spectrum
(\ref{7}) can be obtained if instead of the matrix (\ref{28}) we take the matrix
\begin{equation}
\Lambda_\delta=\frac{1}{1+g^2/4}\left(\begin{array}{cc}1-g^2/4&-g\\ 
g&1-g^2/4\end{array}\right).
\label{Lambda-nonperiodic}
\end{equation}

The most general boundary conditions for the potential $\delta(x)$, that define 
a self-adjoint extension of the Dirac
operator, has  the matrix $\Lambda$ fulfilling  the relation \cite{Benvengu1994}
\begin{equation}
\Lambda\sigma_2\Lambda^\dagger=\sigma_2\,.
\end{equation}
Thus, the matrix $\Lambda$ has four real parameters, for example,
it can be written in the form \cite{Benvengu1994}
\begin{equation}
\Lambda={\rm e}^{{\rm i}\chi}\left( \begin{array}{ll}a&b\\ c&d \end{array}\right),
\label{Lambda-general}
\end{equation}
where $\chi \in [-\pi/2,\, \pi/2]$,  $a,\,b,\,c,\,d \in \R$,   $a d-b c=1$.
The matrices (\ref{7}) and (\ref{Lambda-nonperiodic})
are the representatives of the one-parameter families of the  matrices
\begin{equation}
\Lambda=\left( \begin{array}{ll}\cos\Theta&-\sin\Theta\\ \sin\Theta&
\,\,\,\,\,\cos\Theta \end{array}\right),
\label{Lambda-general2}
\end{equation}
where $\Theta=g$ and $\Theta=2\arctan(g/2)$, respectively.

\subsection{A symmetric double-delta potential: $N=2$, $g_1 =g_2 \equiv -g$,
$r_1 = - r_2 \equiv - R$}
\label{subsec-symmetric-double-delta}

For the potential composed of two identical delta-centers, which are
separated by a distance $l=2R$,
the transmission matrix is the product
$\Lambda = \Lambda_\delta \Lambda_0 \Lambda_\delta$, where the matrix
$\Lambda_\delta$ is defined by Eq.\,(\ref{28}). The matrix $\Lambda_0$ corresponds to
the space between the delta-centers. It is obtained from Eq.\,(\ref{24}),
 replacing $\gamma l$ by $2{\rm i}z$ with $z$ defined as $z=R\sqrt{m^2-E^2}$.
 Therefore the matrix $\Lambda_0$ equals
\begin{equation}
 \Lambda_0 = \left( \begin{array}{ll} \smallskip \cosh(2z)~~~~~~~~ \rho \, \sinh(2z) \\
                      \rho^{-1} \sinh(2z)~~~~~\cosh(2z) \end{array} \right).
\label{32}
\end{equation}
As a result of the matrix multiplication, we obtain the following elements of
the $\Lambda$-matrix:
\begin{equation}
\!\!\!\!\!\!\!\!\!\!\!\!\!\!\!\!\!\!
 \begin{array}{lll} \smallskip
 \lambda_{11} = \lambda_{22} = \cos(2g)\cosh(2z) + {1 \over 2}
 \left( \rho - {1 \over \rho} \right) \sin(2g) \sinh(2z), \\ \smallskip
 \lambda_{12} = - \sin(2g)\cosh(2z) + \left( \rho \cos^2\!g +{1 \over \rho}\sin^2\!g\right)
 \sinh(2z), \\
\lambda_{21} =  \sin(2g)\cosh(2z) +\left( \rho \sin^2\!g + {1 \over \rho}\cos^2\!g\right)
 \sinh(2z). \end{array}
 \label{33}
\end{equation}
Inserting  these expressions into  Eq.\,(\ref{26}) and using the explicit formula
for $\rho$ [see Eq.\,(\ref{def:rho})], we get the equation, which can be written
in the form of the quadratic equation with respect to $(E \tan g) /\sqrt{m^2 -E^2}$ as follows
\begin{eqnarray}
&&	\left({E \tan g \over \sqrt{m^2 -E^2}} \right)^{\!2}-\frac{2}{1-{\rm e}^{-4z}}
\left({E \tan g \over \sqrt{m^2 -E^2}} \right) \nonumber\\
&&	+\, \frac{1-{\rm e}^{-4z}\tan^{2}\!g}{1-{\rm e}^{-4z}}=0.
\label{34}
\end{eqnarray}

This equation can be rewritten in a simpler form as
\begin{equation}
\frac{\sqrt{m^2 - E^2}}{E\pm m {\rm e}^{-2R\sqrt{m^2 - E^2}}}=\tan g.
\label{35}
\end{equation}
In the limit as $R \to \infty$, we have $ E^{-1}\sqrt{m^2 - E^2} = \tan g$
and hence the bound state energy for a single delta-potential given by the
formula (\ref{31}). In the limit as $R \to 0$, Eq.\,(\ref{34}) reduces to
$ E^{-1}\sqrt{m^2 - E^2} = \tan(2g)$, that  corresponds again to the energy of  
one center, but with the coupling constant $2g$.
%
%
\begin{figure}[ht]
	\centering
	\includegraphics[width=0.98\columnwidth]{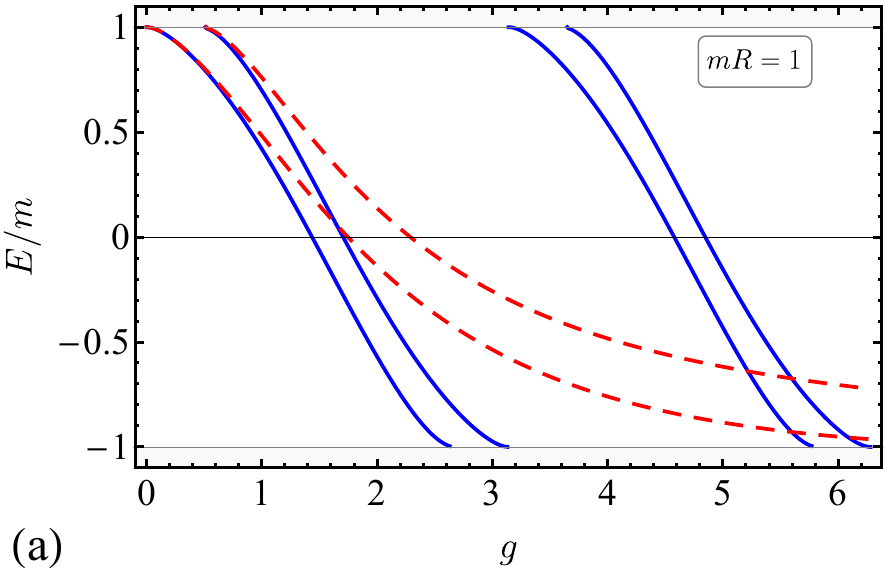}\\
	\includegraphics[width=0.98\columnwidth]{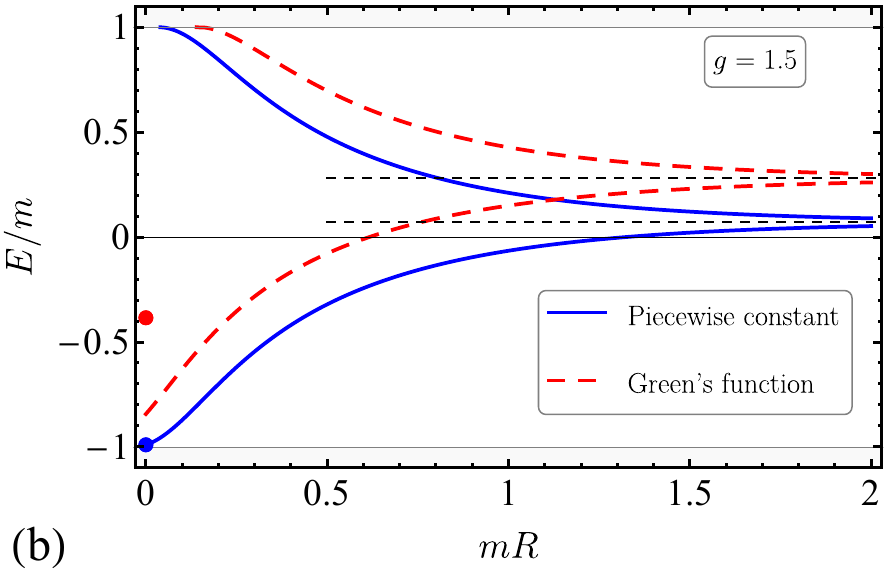}
	\caption{Energy eigenstates in the double-delta potential (a) as functions of coupling constant $g$ at fixed distance $mR=1$; (b) as functions of distance $mR$ for  fixed value $g=1.5$.
	The dashed (red) lines show the energy levels computed by finding the numerical solution to Eq.\,(\ref{12}) in the Green's function approach,
	while the solid (blue) lines -- from Eq.\,(\ref{35}). Dots in panel (b) denote the energy eigenvalues of a
	single delta-potential with  strength $2g$ as given by Eqs.\,(\ref{7}) and (\ref{31}).
	The  dashed (black) lines show the energy eigenvalues in a single delta-potential with strength $g$.
	\label{fig-double}}
\end{figure}

%
%
\begin{figure}[ht]
	\centering
	\includegraphics[width=0.98\columnwidth]{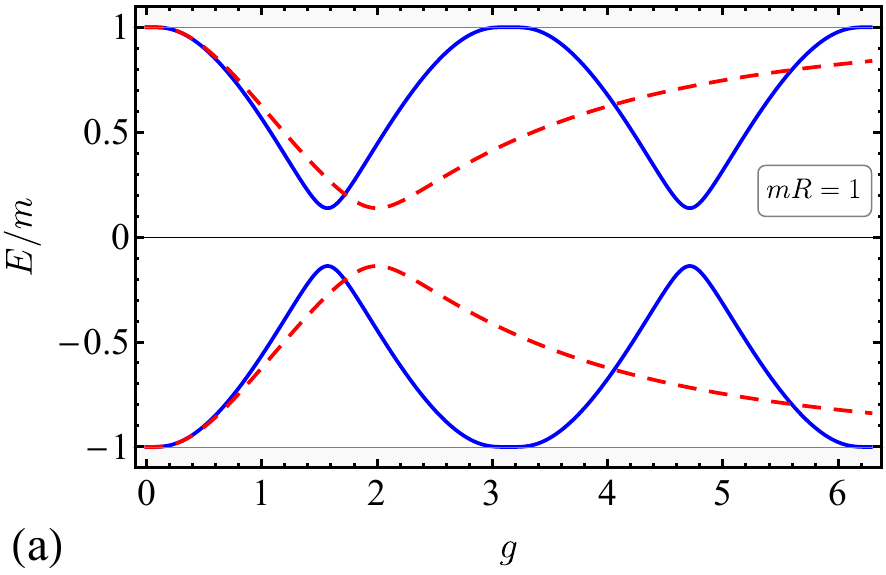}\\
	\includegraphics[width=0.98\columnwidth]{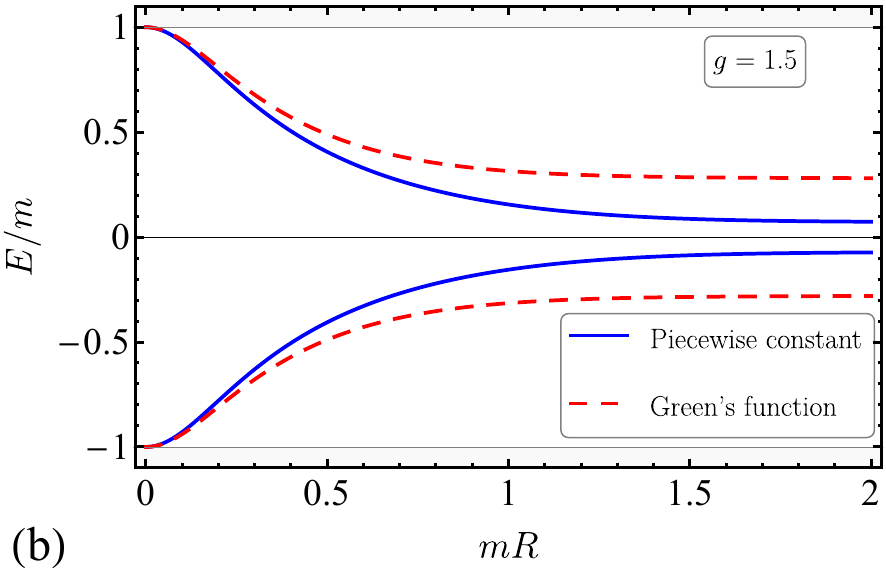}
	\caption{Energy eigenstates in the dipole-like delta-potential (a) as functions of coupling constant $g$ at fixed distance $mR=1$; (b) as functions of distance $mR$ for fixed value $g=1.5$.
	The dashed (red) lines show the energy levels computed by finding the numerical solution to Eq.\,(\ref{20}),
	while the  solid (blue) lines~-- from Eq.\,(\ref{38}).
	\label{fig-dipole}}
\end{figure}

Similarly, one can check that the use of the matrix (\ref{Lambda-nonperiodic}) 
gives the spectrum (\ref{12}).
The energy solutions of Eqs.\,(\ref{12}) and (\ref{35}), as functions of the distance between two equally charged centers,
are shown in Fig.\,\ref{fig-double}(b). At large distances, the energies approach the energy eigenvalues of a single delta-potential given by Eqs.\,(\ref{7}) and (\ref{31}).
For the limit $R\to 0$ as two centers merge at the point $x= 0$, we get
the charge $2g$ for the solution (\ref{35}), whereas the value $2g$ is not continuously reached for the solution (\ref{12})
[the isolated (red) dot corresponds to Eq.\,(\ref{7}) but with $g$ replaced with $2g$].
Moreover, in the configuration space with the use of the matrix 
(\ref{Lambda-nonperiodic}),
we obtain only Eq.\,(\ref{12}) for which the additivity of charges breaks down. 
It would be of interest
to single out a subfamily of the matrices (\ref{Lambda-general}) for which the charge additivity condition is satisfied.

\subsection{A dipole-like delta-potential:  $N=2$, $g_1 = -g_2 \equiv -g$,
$r_1 = -r_2 \equiv -R$ }
\label{subsec-dipole-delta}

For an antisymmetric double-delta potential, the transmission matrix is the
product  $\Lambda =\bar{\Lambda}_\delta\Lambda_0\Lambda_\delta$, where
the matrix $\Lambda_0$ is defined by (\ref{32}) and the matrix $\bar{\Lambda}_\delta$
is obtained from $\Lambda_\delta$, replacing $g$ by $-g$ [see Eq.\,(\ref{28})].
The multiplication of these matrices yields the elements
\begin{equation}
\!\!\!\!\!\!\!\!\!\!\!\!\!
 \begin{array}{llll}
 \smallskip
 \lambda_{11} =  \cosh(2z) + {1 \over 2}
 \left( \rho + {1 \over \rho} \right) \sin(2g) \sinh(2z),  \\
 \smallskip
 \lambda_{12} =   \left( \rho \cos^2\!g -{1 \over  \rho }\sin^2\!g\right)
 \sinh(2z),  \\
 \smallskip
\lambda_{21} =  \left( {1 \over  \rho} \cos^2\!g -\rho \sin^2\!g \right)
 \sinh(2z),  \\
\lambda_{22} =  \cosh(2z) - {1 \over 2}
 \left( \rho + {1 \over \rho} \right) \sin(2g) \sinh(2z). \end{array}
 \label{36}
\end{equation}
Inserting these elements into Eq.\,(\ref{26}) and using the definition 
(\ref{def:rho}), we obtain the equation
\begin{equation}
	\left(\frac{E \tan g}{\sqrt{m^{2}-E^{2}}}\right)^{\!2}=\frac{1
	+ {\rm e}^{-4z}\tan^{2}\!g }	{1 - {\rm e}^{-4z}}\,,
\label{37}
\end{equation}
which can be rewritten as follows
\begin{equation}
 E = \pm \, m \sqrt{\cos^2\!g + {\rm e}^{-4z}\sin^2\!g }\,.
 \label{38}
\end{equation}
This solution also follows from the use of the piecewise constant approximation of
a dipole-like delta-potential \cite{Gorbar}.

In the limit as $R \to 0$, we have the annihilation of the well and the barrier
and obtain the energy of a free fermion: $E = \pm \, m$.
As $R \to \infty$, we get the exact expression for the bound energy of a single
$\delta$-potential given by the formula (\ref{31}).

The use of the matrix (\ref{Lambda-nonperiodic}) gives Eq.\,(\ref{20}).
The behavior of the energy levels (\ref{20}) and (\ref{38})
as functions of the coupling constant and the distance between charges is presented
in Fig.\,\ref{fig-dipole}. The energy levels dependence of
both Eqs.\,(\ref{20}) and (\ref{38}) on the strength $g$ has a non-monotonic behavior.
Moreover, we observe the repulsion of the levels at the values
of $g$, where the levels in the absence of the interaction intersect: $g=2$ 
[see Eq.\,(\ref{18})]
and $g=(n+1/2)\pi$ for Eq.\,(\ref{38}). In the meantime,
the energy levels monotonically depend on the distance $R$ and behave qualitatively
in a similar way as for Eqs.\,(\ref{20}) and (\ref{38}).

We note, however, that in the case of the dipole-like potential with a finite rectangular well
and barrier, the situation is quite different with
respect to the distance dependence of the energy levels. Namely, in this case, for a
supercritical potential strength, the level repulsion is observed
at some distance between the well and barrier. This leads to the interesting phenomenon,
when the wavefunction of the bound state changes its
localization from the negatively charged well to the positively charged barrier as
the distance between the centers is changing \cite{Gorbar}.
This effect, called the migration of a wavefunction, is quite robust and persists to take place in two dimensions for the Coulomb dipole potential \cite{EPL2015} 
(see also the review \cite{FNT2018}).

\subsection{A three-delta potential of the same polarity: $N=3$, $g_1 =g_2=
g_3 \equiv -g$, $r_1 \equiv - R_1$, $r_2 \equiv R_2$, $r_3 =0$}
\label{subsec-symmetric-triple-delta}

Consider the potential composed of  three
identical delta-centers of the same polarity, which are  separated by
distances $R_1$ and $R_2$, respectively.
The $\Lambda$-matrix for this triple-delta potential is the product
$\Lambda =  \Lambda_{\delta}\Lambda_{0,2} 
\Lambda_\delta \Lambda_{0,1}\Lambda_\delta\,,$
where the matrix $\Lambda_\delta$ is defined by Eq.\,(\ref{28}) and
\begin{equation}
 \Lambda_{0,j} = \left( \begin{array}{ll} \smallskip \cosh(z_j)~~~~~~~~ \rho \, \sinh(z_j) \\
                      \rho^{-1} \sinh(z_j)~~~~~\cosh(z_j) \end{array} \right),
\label{39}
\end{equation}
with $z_j := R_j\sqrt{m^2 -E^2}\,, $ $j=1,2$ [compare with  (\ref{32})].
The multiplication of these matrices yields the following expressions for
the $\Lambda$-matrix elements:
\begin{widetext}
\begin{equation}
\begin{array}{llll}
\smallskip
{\lambda_{11} \over \cosh (z_1) \cosh (z_2)}  &=&
\left[ \rho \sin g \, \cos(2g)
-  \rho^{-1} \cos g \, \sin(2g) \right] t_1\, +\, \left[\rho \cos g\, \sin(2g)
-  \rho^{-1}\sin g\,\cos(2g)  \right] t_2 \\
\smallskip
  &+& \cos(3g)\, + \, \left[\tfrac{1}{2}\left( \rho^2 +
\rho^{-2}\right)\sin g\, \sin(2g) + \cos g\,\cos(2g)  \right] t_1t_2\,, \\
\smallskip
{\lambda_{12} \over \cosh (z_1) \cosh (z_2) }& =& -\,
\sin(3g) \,+\,  \left[ \rho \cos g \, \cos(2g)  +   \rho^{-1} \sin g \, 
\sin(2g) \right] (t_1+t_2)\\
\smallskip
&+& \left[\tfrac{1}{2}(\rho-\rho^{-1})^{2} \cos g\,\sin(2g) -
 \rho^{-2} \sin g \right] t_1t_2\,, \\
 \smallskip
{\lambda_{21} \over \cosh (z_1) \cosh (z_2) } &=&   \sin(3g)
\,+\,  \left[ \rho \sin g \, \sin(2g)  +   \rho^{-1} \cos g \, \cos(2g) \right] (t_1+t_2)\\
\smallskip
&-& \left[\tfrac{1}{2}(\rho-\rho^{-1})^{2} \cos g\,\sin(2g) -
	\rho^{2} \sin g \right] t_1t_2\,,\\
\smallskip
{\lambda_{22} \over \cosh (z_1) \cosh (z_2) } &=&
 \left[\rho \cos g\, \sin(2g) -  \rho^{-1}\sin g\,\cos(2g)  \right] t_1\, +\,
 \left[ \rho \sin g \, \cos(2g)   -  \rho^{-1} \cos g \, \sin(2g) \right] t_2 \\
&+& \cos(3g)\, + \,\left[\tfrac{1}{2}\left( \rho^2 +
	\rho^{-2}\right)\sin g\, \sin(2g) + \cos g\,\cos(2g)  \right] t_1t_2\,,
\label{40}
\end{array}
\end{equation}
where $t_j := \tanh (z_j)$,  $j=1,2$.
Inserting the elements (\ref{40}) into  Eq.\,(\ref{26}) and using the relations
\begin{equation}
{1 \over \rho} -\rho = {2E \over \sqrt{m^2 -E^2}}\,,~~
{1 \over \rho^2} + \rho^2 = 2 \left( 1+ {2E^2 \over m^2 -E^2}\right),~~
 \rho^3 -{1 \over \rho^3} =-2 \left[ 3  {E \over \sqrt{m^2 -E^2} }
 + 4\left(  {E \over \sqrt{m^2 -E^2} } \right)^{\!3}\right],
 \label{41}
\end{equation}
we obtain the cubic equation with respect to $(E \tan g)/ \sqrt{m^2 -E^2}$
as follows
\begin{equation}
 \left({E \tan g \over \sqrt{m^2 -E^2}} \right)^{\!3}t_1t_2 \,-\,
 \left({E \tan g \over \sqrt{m^2 -E^2}} \right)^{\!2}(t_1 + t_2 +t_1t_2)
 \,+\, {E \tan g \over \sqrt{m^2 -E^2}} F_1\, +\,F_0 =0,
 \label{42}
\end{equation}
\end{widetext}
where
\begin{equation}
\!\!\!\!\!\!\!\!
 \begin{array}{ll}
 \smallskip
F_0 := \tan^2\!g - {1 \over 4}  \cos^{-2}(g) (1 +t_1)(1 +t_2), \\
F_1 := {1 \over 4} \left[(3 - \tan^2\!g)(1 +t_1 +t_2) + 3  \cos^{-2}(g)\, t_1t_2 \right].
\end{array} \nonumber
\end{equation}

Consider some limiting cases of Eq.\,(\ref{42}). First of these cases is the limit
as $t_2 \to 1$ ($R_2 \to \infty$), fixing the distance $R_1 \equiv R >0$ and setting
$t_1 \equiv t: = \tanh (z)$. Then Eq.\,(\ref{42}) can be rewritten in the form
\begin{widetext}
\begin{equation}
  \left({E \tan g \over \sqrt{m^2 -E^2}} - 1\right) \left\{
 \left({E \tan g \over \sqrt{m^2 -E^2}} \right)^{\!2} -\left( 1 + {1 \over t}\right)
  {E \tan g \over \sqrt{m^2 -E^2}}
  +  {1 \over 2} \left[  1 + {1 \over t} +
\left( 1 - {1 \over t}\right)\tan^2\!g \right] \right\} =0.
 \label{43}
\end{equation}
\end{widetext}
This equation admits the two solutions, one of which describes the bound energy level (\ref{31}) for one delta-center.
The second bracket coincides with Eq.\,(\ref{34}) with a replacement $2z\to z$ and leads thus to the solution of Eq.\,(\ref{35})
derived for the bound state energy of a symmetric double-delta potential.

Another particular case is the limit as all the three centers  merge at the point $x=0$, i.e., $R_1 \to 0$ ($t_1 \to 0$)
and $R_2 \to 0$ ($t_2 \to 0$).  In this case, Eq.\,(\ref{42}) reduces to
\begin{equation}
\!\!\!\!\!\!\!\!\!\!\!
\frac{\sqrt{m^2 - E^2}}{E}= \frac{ (3 \cos^2\!g - \sin^2\!g )\tan g}{1 - 4\sin^2\!g}=\tan(3g),
\label{44}
\end{equation}
from which we find the bound state energy
\begin{equation}
E=   m\, \cos(3g)\,{\rm sign}(\sin(3g)),
\label{45}
\end{equation}
that corresponds to a single delta-potential
given by the formula (\ref{31}), where the coupling constant $g$ is replaced by $3g$.

Finally, consider the limit as $R_1 \to \infty$ and $R_2 \to \infty $
($t_1 \to 1$ and $t_2 \to 1$). In this case, Eq.\,(\ref{42}) reduces to
$ E \tan g = \sqrt{m^2 - E^2}\,,$
from which we immediately get the solution (\ref{31}) for the energy level
of  one delta-center.

%
%
%
\begin{figure}[ht]
	\centering
	\includegraphics[width=0.98\columnwidth]{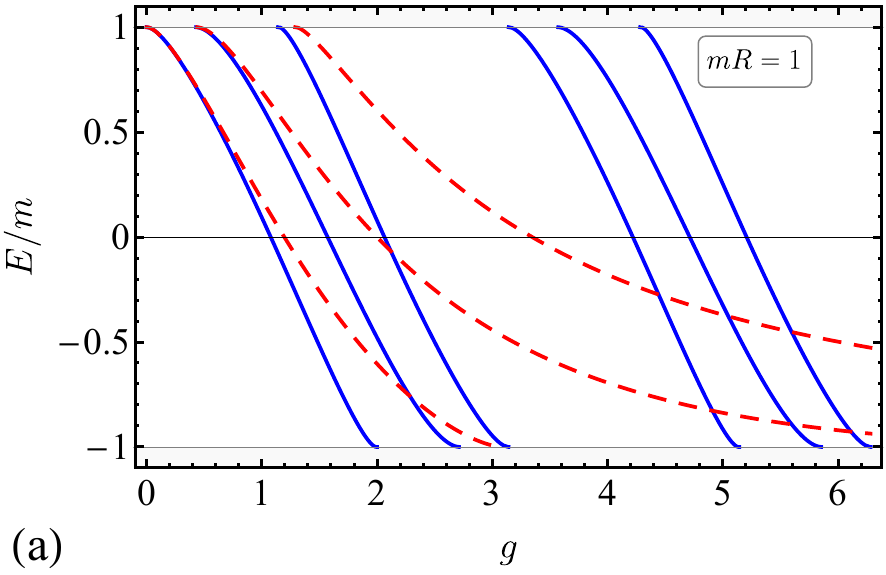}\\
	\includegraphics[width=0.98\columnwidth]{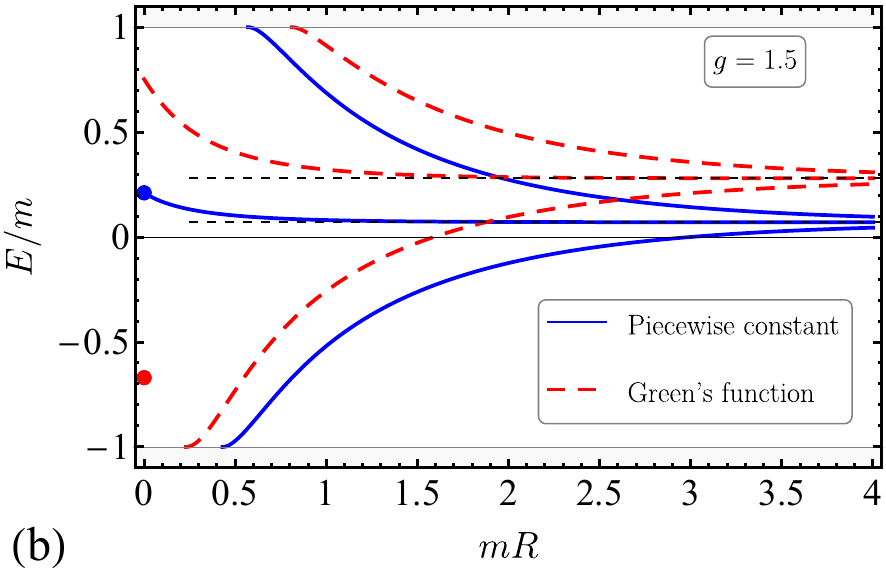}
	\caption{Energy eigenstates of the symmetric triple-delta potential ($g_1=g_2=g_3 \equiv - g$,
		$R_1=R_2 \equiv R$) (a) as functions of coupling constant $g$ at fixed distances $mR=1$;
		(b) as functions of distance $mR$ for fixed value $g=1.5$. The solid (blue) lines show
			different solutions of Eq.\,(\ref{43}), while the dashed (red) lines give the solutions of Eq.\,(\ref{eq-3s-X}) in the Green's function approach. 
			The blue and red dots in panel (b) denote the energy eigenvalues of
			a single delta-potential with the strength $3g$ as given by Eqs.\,(\ref{31}) and (\ref{7}), respectively.
			The dashed (black) lines show the energy eigenvalues of a single delta-potential with strength $g$.
		\label{fig-triple-plus}}
\end{figure}

Note that Eq.\,(\ref{42}) can be rewritten through  the variables $z_1$ and $z_2$
as follows
\begin{eqnarray}
&& G_3 \left({E \tan g \over \sqrt{m^2 -E^2}} \right)^{\!3} \,+\,
G_2 \left({E \tan g \over \sqrt{m^2 -E^2}} \right)^{\!2} \nonumber\\
&&+ \,\, G_1 {E \tan g \over \sqrt{m^2 -E^2}} \, +\,G_0=0,
 \label{46}
\end{eqnarray}
where
\begin{equation}
\begin{array}{llll}
\smallskip
G_0 := \left[ {\rm e}^{-2z_1} + {\rm e}^{-2z_2}+ {\rm e}^{-2(z_1 +z_2)}
\right] \tan^2\!g -1, \\
\smallskip
G_1 := 3 - \left[ {\rm e}^{-2z_1} + {\rm e}^{-2z_2}-
{\rm e}^{-2 (z_1 +z_2)} \right] \tan^2\!g ,\\
\smallskip
G_2 := \left( 1+ {\rm e}^{-2z_1}\right) \left(1 + {\rm e}^{-2z_2} \right)-4,\\
 G_3: = \left( 1- {\rm e}^{-2z_1}\right) \left(1 - {\rm e}^{-2z_2} \right).
\end{array}
\nonumber
\end{equation}

Figure~\ref{fig-triple-plus} shows the energy eigenvalues of a symmetric triple-delta potential with equal distances between the delta-functions. Panel (a) describes their behavior with the change of the coupling constant $g$.
Three levels emerge from the upper continuum one by one, monotonically shift down and, finally, dive into the lower
continuum. While the last level disappears at $g=\pi$, the new one emerges from the upper continuum and the whole process
is repeating. Figure~\ref{fig-triple-plus}(b) shows the energy levels for a fixed value of $g$ and for different distances
between the delta-functions. If the distance is very small ($R\to 0$), there is only one energy level, which approaches
the eigenvalue in a single delta-potential with the strength $3g$ given by 
Eq.\,(\ref{45}) (this value is depicted by
the blue dot). Then, with an increase of the distance, the two new levels emerge from the upper continuum and, finally,
as $R\to\infty$, all the three of them tend to the eigenvalue of a single delta-potential with the strength $g$ (shown by
the dashed  black line).

\subsection{A three-delta potential with the middle center of opposite polarity:
$N=3$, $g_1 =g_2= - g_3 \equiv - g$, $r_1 \equiv - R_1$, $r_2 \equiv R_2$, $r_3 =0$}
\label{subsec-alternating-triple-delta}

Consider now the triple-delta potential with a middle delta-center and
the corresponding lateral centers being  of opposite polarity.
The $\Lambda$-matrix in this case is the product
$\Lambda =\Lambda_{\delta}\Lambda_{0,2} \bar{\Lambda}_\delta \Lambda_{0,1}\Lambda_\delta,$
where the matrices $\Lambda_\delta$ and $\Lambda_{0,j}$, $j=1,2$, are given by
Eqs.\,(\ref{28}) and (\ref{39}), respectively. As above, the matrix $\bar{\Lambda}_\delta$
is obtained from $\Lambda_\delta$ replacing $g$ by $-g$.
The multiplication of these matrices leads to the following elements
of the $\Lambda$-matrix:
\begin{widetext}
\begin{equation}
 \begin{array}{llll}
 \smallskip
{\lambda_{11} \over \cosh (z_1) \cosh (z_2)} = \cos g + \rho \sin g \, t_1
- \rho^{-1} \sin g\,t_2 -\left[\tfrac{1}{2}\left( \rho^2 +\rho^{-2}\right)\sin g\, \sin(2g)
- \cos g\,\cos(2g)  \right] t_1t_2 \\
\smallskip
{ \lambda_{12} \over \cosh (z_1) \cosh (z_2) }  = -  \sin g + \rho \,\cos g\,( t_1 +
t_2) -\left[\tfrac{1}{2}\left( \rho +\rho^{-1}\right)^{2}\cos g\, \sin(2g)
- \rho^{-2}\sin g \right] t_1 t_2  \,, \\
\smallskip
{ \lambda_{21} \over \cosh (z_1) \cosh (z_2) }  = \sin g + \rho^{-1} \,\cos g\,( t_1 +
t_2) + \left[\tfrac{1}{2}\left( \rho +\rho^{-1}\right)^{2}\cos g\, \sin(2g)
- \rho^{2}\sin g \right] t_1 t_2  \,, \\
{\lambda_{22} \over \cosh (z_1) \cosh (z_2) } = \cos g - \rho^{-1} \sin g \, t_1
 +  \rho \sin g\,t_2 -\left[\tfrac{1}{2}\left( \rho^2 +\rho^{-2}\right)\sin g\, \sin(2g)
- \cos g\,\cos(2g)  \right] t_1t_2\,.
 \end{array}
\label{47}
\end{equation}
\end{widetext}
Inserting these elements into Eq.\,(\ref{26}) and using the relations (\ref{41}),
we obtain the equation, which is similar to (\ref{42}), again having the cubic
form with respect to $(E \tan g)/ \sqrt{m^2 -E^2}$\,:
\begin{equation}
\!\!\!\!\!\!\!\!\!\!\!\!\!\!
\left({E \tan g\over \sqrt{m^2 -E^2}} -1 \right)\!
\left[ \left({E\tan g \over \sqrt{m^2 -E^2} }\right)^{\!2}t_1t_2 -D \right]  =0,
\label{48}
\end{equation}
where
\begin{equation}
D := { 1 \over 4}\cos^{-2}\!g \,\, (1 +t_1)(1 +t_2) -\tan^2\!g  \, t_1t_2\, . 
\nonumber
\end{equation}

One of the solutions to Eq.\,(\ref{48}) is the bound state energy (\ref{31})
of a single-delta potential. This state exists for any finite
configuration of the lateral delta-centers. The other two solutions are found from
the equation
\begin{equation}
{E^2 \over m^2 -E^2} = {1 \over 4 \sin^2\!g }\left( 1 + {1 \over t_1}\right)
\left( 1 + {1 \over t_2}\right) -1 ,
\label{49}
\end{equation}
which can be rewritten  in the form
\begin{equation}
E = \pm\, m \sqrt{1 - \left( 1 - {\rm e}^{-2z_1}\right)
\left( 1 - {\rm e}^{-2z_2}\right) \sin^2\!g}\,.
\label{50}
\end{equation}
In the limit as $R_1 \to \infty$ ($z_1 \to \infty$), $R_2 \equiv R$,
this equation reduces to (\ref{38}) with an exchange $2z\to z$,
describing the bound state energy of a dipole delta-potential. The annihilation
of the delta-centers of opposite polarity occurs if
$R_1 \to 0$ or $R_2 \to 0$, resulting in the appearance of a free fermion: $E = \pm m$.

\begin{figure}[ht]
	\centering
	\includegraphics[width=0.98\columnwidth]{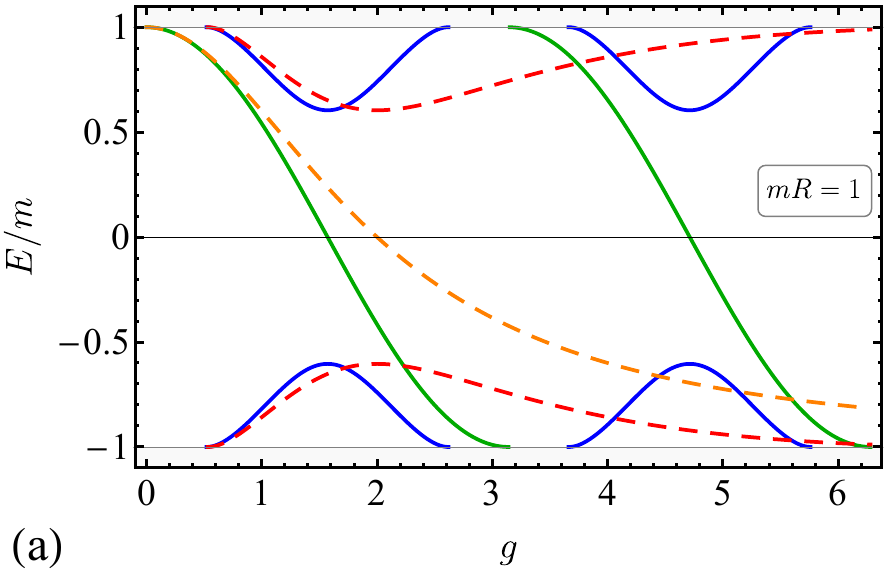}\\
	\includegraphics[width=0.98\columnwidth]{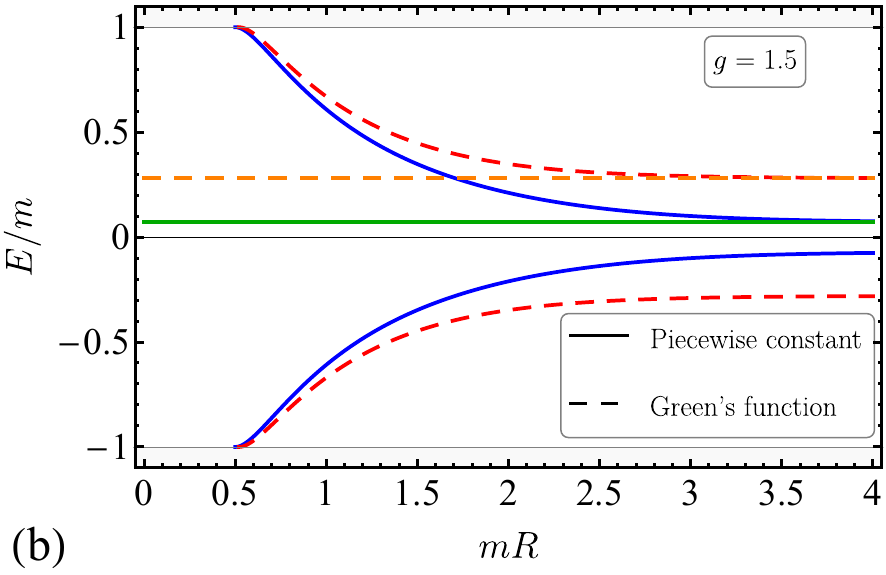}
	\caption{Energy eigenstates of the alternating triple-delta potential
		($g_1=g_2=-g_3 \equiv - g$, $R_1=R_2 \equiv R$) (a) as functions of coupling constant $g$ at fixed distances
		$mR=1$; (b) as functions of distance $mR$ for fixed value $g=1.5$.
		The solid (blue and green) lines show different solutions of Eq.\,(\ref{48}): the green line corresponds to zero value of the first bracket, while the blue lines show the solutions of Eq.\,(\ref{50}). The dashed (red and orange) lines depict different solutions of Eq.\,(\ref{eq-3a-X}) in the Green's function approach: the orange line corresponds to zero value of the first bracket, while the red lines give the solutions of Eq.\,(\ref{energy-3a}).
		\label{fig-triple-alt}}
\end{figure}

The behavior of energy levels for a symmetric configuration ($R_{1}=R_{2}$) is shown
in Fig.\,\ref{fig-triple-alt}. In panel (a) they are plotted as functions of $g$ for fixed distances between
the delta-centers, while in  panel (b)  as functions of the distance for a fixed coupling constant.
First of all, it is clearly seen that the spectrum consists of two separate components: one energy level for
a single delta-potential with the strength $g$ (it is shown by the green line in both panels) and two levels
in the dipole delta-potential (blue lines). The green line does not depend on the distance between centers
meaning that it is indeed decoupled from the rest of the system. The blue lines are located symmetrically
with respect to zero energy, which is a manifestation of the charge conjugation symmetry of the dipole potential.
Therefore, we see that the triple-delta potential with alternating signs is the superposition of the cases considered
in Sec.\,\ref{subsec-single-delta} and Sec.\,\ref{subsec-dipole-delta}.

\bigskip

\section{Concluding remarks}

In the present work,  we have examined the one-dimensional Dirac equation with the potential
composed of one, two, and three delta-function centers. The corresponding models are
exactly solvable and therefore all the results for bound state energies of this equation
are given in closed expressions through transcendental equations. These equations
were solved numerically, as illustrated in Figs.\,\ref{fig-double}--
\ref{fig-triple-alt}, where the dependence
of bound state energies  both on the strength of interaction and on the inter-center
distance is clearly demonstrated.

For finding the bound state energies, we have utilized two different approaches. 
One of them uses the Green’s function method. 
The other one, similarly to the procedure used in \cite{Calkin},
starts from a piecewise constant approximation of one delta-function,
yielding in a squeezing limit the matrix that connects  boundary conditions.
Having such a matrix for one delta-center, the total transmission matrix
for a multiple delta-function potential is found just multiplying 
the one-center connection 
matrices and the free transfer matrices between neighbor centers. We note that the results
obtained by the Green's function method are reproduced within the second approach 
if one uses the connection 
matrix (\ref{Lambda-nonperiodic}). At the same time, two types of equations obtained in framework of the second
approach by means of the connection matrices (\ref{28}) and  (\ref{Lambda-nonperiodic}) are related through 
the transformation mentioned after Eq.\,(\ref{Lambda-general2}).

As discovered in \cite{sm}, even in the case of a single delta-potential,
different methods yield different boundary conditions. Hence, it is not surprise
to have ambiguous results for multiple delta-function potentials. However,
due to the presence of several delta-centers in the total potential,
it was possible to check the principle of strength additivity in the limit cases
as the inter-center distance converges to zero or infinity. Verifying this principle
in each model, one can argue in favour of its application. Otherwise,
because of the ambiguity problem, the one-dimensional Dirac equation with
delta-like potentials  does not make a physical sense if it is considered alone
without any additional conditions describing the way how a given delta-center is
defined. In this regard, the piecewise constant approximation of
delta-potentials seems to be a physically motivated approach because of
satisfying the principle of strength additivity.

In general, similarly to the
studies \cite{c-g,Golovaty}, one can conclude that the Dirac equation with
point-like potentials contains hidden parameters, which are responsible
for the ``pathway'' of materializing point interactions.

In conclusion, it would be interesting to generalize the results of the present paper  to more realistic potentials,
for example, to study the one-dimentional Dirac equation with two charged centers interacting through the Coulomb potential.
The recent paper \cite{Loran2022} provides the treatment of the scattering problem for the Schr\"{o}dinger equation
with multi-delta-function potentials
 in two and three dimensions. The generalization of our results for the Dirac equation with multiple
delta-potentials to higher dimensions is also of great interest.

\bigskip
{\bf  Acknowledgments}\\

\smallskip
It is a pleasure to dedicate this research to Alexander S. Davydov on the occasion of
his 110th birthday and  thank him for the collaboration, support and encouragement
of one of us (A.V.Z.) over many years. V.P.G. thanks Dr. F.~Fillion-Gourdeau for the correspondence
and discussions on double delta potentials. V.P.G., A.V.Z and Y.Z. 
acknowledge the partial financial
support from the National Academy of Sciences of Ukraine, Project No.~0122U002313.
We would like to thank the Armed Forces of Ukraine
for providing security to perform this work.

\bigskip
{\bf  References}\\


\begin{thebibliography}{9}


\bibitem{do}
Y.N.\,\,Demkov and V.N.\,\,Ostrovskii, {\it Zero-Range Potentials and
Their Applications in Atomic Physics}, Leningrad University Press (1975)
[Plenum Press, NY (1988)].

\bibitem{a-h}
 S.\,\,Albeverio,  F.\,\,Gesztesy, R.\,\,H{\o}egh-Krohn, and H.\,\,Holden,
 {\it Solvable Models in Quantum Mechanics (With  an Appendix Written by Pavel Exner).
 2nd revised edn}, AMS Chelsea Publishing, Providence, RI (2005).

\bibitem{sm}
B.\,\,Sutherland and D.C.\,\,Mattis, {\it Phys. Rev.~A} {\bf 24},  1194 (1981).

\bibitem{c-g}
P.L.\,\,Christiansen, H.C.\,\,Arnbak, A.V.\,\,Zolotaryuk, V.N.\,\,Ermakov,
 and Y.B.\,\,Gaididei,  {\it J.~Phys.~A: Math. Gen.} {\bf 36}, 7589 (2003);
A.V.\,\,Zolotaryuk, P.L.\,\,Christiansen, and S.V.\,\,Iermakova,
{\it J.~Phys.~A: Math. Gen.} {\bf 39}, 9329 (2006);
F.M.\,\,Toyama  and Y.\,\,Nogami,  {\it J.~Phys.~A: Math. Theor.} {\bf 40}, F685 (2007).

 \bibitem{Golovaty}
   Y.D.\,\,Golovaty and S.S.\,\,Man'ko, {\it Ukr. Math. Bull.} {\bf 6}, 169 (2009);
 Y.D.\,\,Golovaty and  R.O.\,\,Hryniv, {\it J.~Phys.~A: Math. Theor.}
   {\bf 43}, 155204 (2010), {\bf 44}, 049802 (2011);  {\it Proc. Royal Soc. Edinb.}
  {\bf 143A}, 791 (2013); Y.\,\,Golovaty, {\it Integr. Equ. Oper. Theory}
  {\bf 75}, 341 (2013).

\bibitem{Subramanian1971}R.~Subramanian and K.V.~Bhagwat, Phys. Stat. Sol. B {\bf48}, 399 (1971).

\bibitem{Lapidus}
I.R.\,\,Lapidus, {\it Am.~J.~Phys.} {\bf 51}, 1036 (1983).

\bibitem{Calkin}
M.G.\,\,Calkin, D.\,\,Kiang, and Y.\,\,Nogami, {\it Am.~J.~Phys.} {\bf 55}, 737 (1987).

\bibitem{Benvengu1994}
S.\,\,Benvegnu and L.\,\,D\c{a}browski, {\it Lett.~Math.~Phys.} {\bf 30}, 159 (1994).

\bibitem{hu}
R.J.\,\,Hughes, {\it Lett.~Math.~Phys.} {\bf 34}, 395 (1995);
{\it Rep.~Math.~Phys.}  {\bf 39}, 425 (1997).

 \bibitem{bcl}
R.D.\,\,Benguria, H.\,\,Castillo, and M.\,\,Loewe,  {\it J.~Phys.~A: Math. Gen.}
{\bf 33}, 5315 (2000).

\bibitem{adv}
V.\,\,Alonso and S.\,\,De~Vincenzo, {\it Int. J.~Theor. Phys.} {\bf 39}, 1483 (2000).

 \bibitem{km}
H.J.\,\,Korsch and S.\,\,Mossmann, {\it J.~Phys.~A: Math. Gen.} {\bf 36}, 2139 (2003).

\bibitem{ntd}
Y.\,\,Nogami, F.M.\,\,Toyama, and van\,\,W.\,\,Dijk, {\it Am.~J.~ Phys.}
{\bf 71}, 950 (2003).

\bibitem{Peter}
H.\,\,Arnbak, P.L.\,\,Christiansen, and Y.B.\,\,Gaididei, {\it Phil. Trans.~R.~Soc.~A}
 {\bf 369}, 1228 (2011).

\bibitem{Guilarte}
J.M.\,\,Guilarte, J.M.\,\,Munoz-Castaneda, I.\,\,Pirozhenko, and
L.\,\,Santamaría-Sanz, {\it Front. Phys.}  {\bf 7}, 109 (2019).

 \bibitem{Fillion}
F.\,\,Fillion-Gourdeau, E.\,\,Lorin, and A.D.\,\,Bandrauk, {\it J.~Phys.~A:
Math. Theor.} {\bf 45}, 215304 (2012).

\bibitem{Gorbar}
E.V.\,\,Gorbar,  V.P.\,\,Gusynin,  and O.O.\,\,Sobol, {\it Phys.~Rev.~B} {\bf 92},
235417 (2015).

\bibitem{cs}
T.\,\,Cheon and T.\,\,Shigehara, {\it Phys.~Lett.~A} {\bf 243}, 111 (1998);
P.\,\,Exner, H.\,\,Neidhardt, and V.A.\,\,Zagrebnov, {\it Commun. Math. Phys.}
{\bf 224}, 593 (2001).

\bibitem{Erman}
F.\,\,Erman, M.\,\,Gadella, S.\,\,Tunali, and H.\,\,Uncu, {\it Eur. Phys.~J.~Plus} {\bf 132}, 352 (2017).

\bibitem{zz14}
 A.V.\,\,Zolotaryuk and Y.\,\,Zolotaryuk,
 {\it Int.~J.~Mod. Phys.~B}   {\bf 28}, 1350203 (2014);
 {\it Low Temp. Phys.}   {\bf 46}, 927 (2020).

\bibitem{mcks87prc}
B.H.J.\,\,McKellar, and G.J.\,\,Stephenson Jr., {\it Phys. Rev. C} {\bf 35},  
2262 (1987).

\bibitem{EPL2015}
E.V.\,\,Gorbar, V.P.\,\,Gusynin, and O.O.\,\,Sobol, {\it Europhys. Lett.} {\bf 111},  37003 (2015).

\bibitem{FNT2018}
E.V.\,\, Gorbar, V.P.\,\,Gusynin, and O.O.\,\,Sobol, {\it Low Temp. Phys.}   {\bf 44}, 491 (2018).

\bibitem{Loran2022}
F.\,\,Loran and A.\,\,Mostafazadeh, {\it Ann. Phys. (NY)}  {\bf 443}, 168966 (2022).

\end{thebibliography}
\end{document}